# Balanced Allocation on Graphs

Krishnaram Kenthapadi[*]  Rina Panigrahy[†]

## Abstract

It is well known that if $n$ balls are inserted into $n$ bins, with high probability, the bin with maximum load contains $(1 + o(1)) \log n / \log \log n$ balls. Azar, Broder, Karlin, and Upfal [1] showed that instead of choosing one bin, if $d \geq 2$ bins are chosen at random and the ball inserted into the least loaded of the $d$ bins, the maximum load reduces drastically to $\log \log n / \log d + O(1)$. In this paper, we study the two choice balls and bins process when balls are not allowed to choose any two random bins, but only bins that are connected by an edge in an underlying graph. We show that for $n$ balls and $n$ bins, if the graph is almost regular with degree $n^{\epsilon}$, where $\epsilon$ is not too small, the previous bounds on the maximum load continue to hold. Precisely, the maximum load is $\log \log n + O(1/\epsilon) + O(1)$. So even if the graph has degree $n^{\Omega(1/\log \log n)}$, the maximum load is $O(\log \log n)$. For general $\Delta$-regular graphs, we show that the maximum load is $\log \log n + O(\frac{\log n}{\log(\Delta/\log^4 n)}) + O(1)$ and also provide an almost matching lower bound of $\log \log n + \frac{\log n}{\log(\Delta \log n)}$. Further this does not hold for non-regular graphs even if the minimum degree is high.

Vöcking [29] showed that the maximum bin size with $d$ choice load balancing can be further improved to $O(\log \log n/d)$ by breaking ties to the left. This requires $d$ random bin choices. We show that such bounds can be achieved by making only two random accesses and querying $d/2$ contiguous bins in each access. By grouping a sequence of $n$ bins into $2n/d$ groups, each of $d/2$ consecutive bins, if each ball chooses two groups at random and inserts the new ball into the least-loaded bin in the lesser loaded group, then the maximum load is $O(\log \log n/d)$ with high probability. Furthermore, it also turns out that this partitioning into aligned groups of size $d/2$ is also essential in achieving this bound, that is, instead of choosing two aligned groups, if we simply choose random but possibly unaligned random sets of $d/2$ consecutive bins, then the maximum load jumps to $\Omega(\log \log n/ \log d)$ even if the two sets are always chosen to be disjoint.

## 1   Introduction

The analysis of balls and bins has several interesting applications including hashing and online load balancing. It is well known that if $n$ balls are randomly thrown into $n$ bins, with high probability, the bin with maximum load contains $(1 + o(1))\frac{\log n}{\log \log n}$ balls [19]. Azar, Broder, Karlin, and Upfal [1] showed that instead of choosing one bin, if $d \geq 2$ bins are chosen at random, the maximum load reduces drastically to $\frac{\log \log n}{\log d} + O(1)$. Vöcking [29] showed that asymmetry helps in load balancing; when bins have equal size, if ties are broken to the left, then the maximum bin size drops to $\frac{\log \log n}{d \ln \phi_d} + O(1)$ where $\phi_d$ is a constant $> 1$ that approaches $\ln 2$ for large $d$. He also showed that the above bound is tight, that is, the maximum load is at least $\frac{\log \log n}{d \ln \phi_d} - O(1)$ even if the $d$ bins are chosen from an arbitrary fixed distribution, and irrespective of the policy used to decide the bin out of the $d$ choices for a given ball. Berenbrink et al. [3] extended these results to the case when the number of balls $m$ is greater than the number of bins $n$ showing that the maximum height is at most $\frac{\log \log n}{\ln d} + O(1)$ above the average when $d$ bins are chosen at random and similar results when ties are broken asymmetrically. The maximum load can also be reduced by moving previously inserted balls into their alternate bin choice(s) when a new ball is inserted [24, 18]; with two choice load balancing, by performing at most $h$ moves per insert, we can maintain a maximum load of $O(\frac{\log \log n}{h \log(\log \log n/h)})$ for $n$ balls and $n$ bins [25]. Further there has been a lot of work on parallel balls and bins where balls are thrown parallelly into bins in rounds [2, 7, 13].

A natural extension of the earlier work is to consider a graph-based model where balls are not allowed to choose any two random bins but only bins that are connected by an edge in a given underlying graph. While the earlier studies use a complete underlying graph, an understanding of balls and bins processes over arbitrary graphs is not only theoretically interesting but also meaningful in pratical scenarios. For example, if bins are arranged in a line, what if balls are only allowed to choose two random bins that are close-by, say at most distance $\Delta$ apart – here the underlying graph has edges only when the bins are at most distance $\Delta$ apart. How small can $\Delta$ be so that the earlier results continue to hold? This is useful from a practical point of view because the cost of accessing the two bins may depend on the distance between them. For example, if the bins represent hash buckets of a hash table stored on disk, then after having accessed the first bucket, the time required to access the second bucket depends on how far the head has to move from the first to the second bucket, as the access time is dominated by the seek time. Or, it may be easier to read memory in large bursts and accomplish the read of the two buckets that are at most $\Delta$ apart in one random access. Similarly when $d$ random bins are explored, it may be beneficial to group the accesses into as few bursts as possible.

[*]Department of Computer Science, Stanford University, Stanford, CA 94305. Supported in part by NSF Grants EIA-0137761 and ITR-0331640, and a grant from SNRC. kngk@cs.stanford.edu.

[†]Department of Computer Science, Stanford University, Stanford, CA 94305. Supported in part by Stanford Graduate Fellowship, NSF Grants EIA-0137761 and ITR-0331640, and a grant from SNRC. rinap@cs.stanford.edu.

Another example could be the load-balancing of requests in a network setting. To apply the two-choice heuristic, a client may make a request to a server. The server may then query another random server and the client may be serviced by the least-loaded of the two servers. However it may be inefficient for the server to query any other server; perhaps it only knows about a few servers, or it is close to only a few servers, or only a few other servers may be able to service the client's request. So instead it may be connected to $\Delta$ neighboring servers and could only query one of these at random. This amounts to the following question: If each ball picks a random bin and that bin again selects a random "neighboring" bin and the ball is inserted into the least-loaded of the two bins, then what is the maximum load? This is equivalent to picking a random edge in the graph for each ball. For what graphs, do the results for two choices continue to hold? First we show a lower bound of $\Omega(\log \log n + \frac{\log n}{\log(\Delta \log n)})$ for maximum bin size where $\Delta$ is the average degree of the graph; in particular if the average degree is $n^\epsilon$ for some constant $\epsilon$, then the lower bound implies a maximum load of $\Omega(\log \log n + 1/\epsilon)$. However, if the graph is not regular, we show that the maximum bin size can be much higher; even for graphs with minimum degree $n^\epsilon$, it can be $\Omega(\log n/\log \log n)$. On the other hand, for $\Delta$-regular graphs, the maximum bin size is $\log \log n + O(\frac{\log n}{\log(\Delta/\log^4 n)}) + O(1)$, which almost matches the lower bound (Section 2). This means that the earlier bound on the maximum load with two choice load balancing (on a complete graph) continues to hold on almost regular graphs with degree as low as $n^\epsilon$ for any constant $\epsilon$ – the maximum load is $\log \log n + O(1/\epsilon) + O(1)$. So even if the graph has degree only $n^{\Omega(1/\log \log n)}$, the maximum load is still $O(\log \log n)$. We also show that when the number of balls $m$ is more than the number of bins $n$, the earlier results by Berenbrink et al. [3] do not carry over; even for graphs of degree $n^\epsilon$ where $\epsilon$ is constant, the maximum load can be $m/n + \omega(\log \log n)$. We achieve similar results when moves are allowed (Section 3).

Further, Vöcking's observation that the maximum load of $O(\frac{\log \log n}{d})$ can be achieved by breaking ties asymmetrically requires $d$ random bin choices. Is it possible to achieve similar bounds when the $d$ choices may not be random, but preferably close together or in a few groups where bins within a group are close-by? Such an access pattern may be more efficient especially when bins represent memory locations in a hash table. We show that such bounds can be achieved by making two random accesses and querying $d/2$ contiguous bins in each access, even if ties are not broken asymmetrically. By partitioning a sequence of $n$ bins into $2n/d$ groups, each of $d/2$ consecutive bins, if each ball chooses two groups at random and inserts the new ball into the least-loaded bin in the lesser loaded group, then the maximum load is $2 \log \log n/d + O(1)$ with high probability (Section 4). This is surprising as it shows that querying $d$ bins in two groups of size $d/2$ outperforms querying $d$ random bins! Furthermore, it also turns out that this partitioning into aligned groups of size $d/2$ is also essential in achieving this bound. Instead of choosing two aligned groups, if we simply choose two disjoint but possibly unaligned random sets of $d/2$ consecutive bins – which only seems like a simple, natural variation – then the maximum load jumps to $\frac{\log \log n}{\log d}$ ! Similarly we also show that it is important to insert into the lesser loaded group. In fact, we can get arbitrarily close to Vöcking's bound of $\frac{\log \log n}{d \ln \phi_d} + O(1)$ by using $c$ random groups of size $d/c$ each and breaking ties asymmetrically. Moreover, our result implies that we can achieve constant load by making two burst accesses of $\log \log n$ bins each, provided these accesses are made from a set of disjoint groups; otherwise the maximum bin size becomes $\frac{\log \log n}{\log \log \log n}$.

## 2 Two choice Load Balancing on $\Delta$-regular Graphs

One method to obtain bounds on maximum load with two choice load balancing is to use the layered induction technique [1] that recursively bounds the fraction of bins, $p_i$ with load $i$. For a new ball to fall at a height of $i+1$ (that is, into a bin with at least $i$ balls), both of its bin choices must have at least $i$ balls, which happens with probability at most $p_i^2$. This essentially implies that $p_{i+1} \approx p_i^2$, implying a quadratic drop with height, giving a maximum load of $\log \log n$ with high probability. This technique fails on arbitrary graphs, because the probability that both the bin choices for a subsequent ball insert have load at least $i$ is no longer $p_i^2$ as the two bins are not chosen independently and randomly. Alternatively the witness tree method tracks the occurrence of a bin with a high load using a suitable tree of events and shows that such a tree is unlikely to occur. This type of analysis was introduced in the context of PRAM simulations [7, 22] and later adapted for balls and bins [6, 5, 29]. Using this method, we analyze the maximum load when the underlying graph is $\Delta$-regular.

First we prove a lower bound on the maximum load for a $\Delta$-regular graph.

**Theorem 2.1** *Given a $\Delta$-regular graph with $n$ nodes representing $n$ bins, if $n$ balls are thrown into the bins by choosing a random edge and placing into the smaller of the two bins connected by the edge, then the maximum load is at least $\Omega(\log \log n + \frac{\log n}{\log(\Delta \log n)})$ with high probability of $1 - 1/n^{\Omega(1)}$.*

**Proof**: The $\Omega(\log \log n)$ term in the lower bound follows from Vöcking's lower bound [29] for placement of $n$ balls into $n$ bins if each ball picks two bins at random using any arbitrary but fixed distribution. The second term follows by analyzing the ratio of the number of balls to the number of edges: since there are $n$ balls and $n\Delta/2$ edges, with high probability, some edge will get $\frac{\log n}{\log(\Delta \log n/2)}$ balls (this is like throwing $n$ balls into $n\Delta/2$ bins). Hence at least one of the endpoints of this edge will get $\frac{\log n}{2 \log(\Delta \log n/2)}$ balls. □

This implies that if the bins are arranged in a line and if each ball chooses two bins that are at most distance $\Delta$ apart, the maximum load is at least $\Omega(\log \log n + \frac{\log n}{\log(\Delta \log n)})$.[1] In particular, when $\Delta$ is $polylog(n)$ the maximum bin load is $\Omega(\frac{\log n}{\log \log n})$. If $\Delta = n^\epsilon$ for some constant $\epsilon$, then the lower bound implies a maximum load of $\Omega(\log \log n + 1/\epsilon)$.

The following theorem gives an almost matching upper bound on the maximum load for $\Delta$-regular graphs.

**Theorem 2.2** *Given a $\Delta$-regular graph with $n$ nodes representing $n$ bins, if $n$ balls are thrown into the bins by choosing a random edge and placing into the smaller of the two bins connected by the edge, then the maximum load is $\log \log n + O(\frac{\log n}{\log(\Delta/\log^4 n)}) + O(1)$ with high probability of $1 - 1/n^{\Omega(1)}$.*

This holds even if the graph is almost regular, that is, each node has degree, $\Theta(\Delta)$.

**Corollary 2.1** *Given a $n^\epsilon$-regular graph with $n$ nodes representing $n$ bins, if $n$ balls are thrown into the bins by choosing a random edge and placing into the smaller of the two bins connected by the edge, then for any $\epsilon > 8 \log \log n / \log n$, the maximum load is $\log \log n + O(1/\epsilon) + O(1)$ with high probability of $1 - 1/n^{\Omega(1)}$.*

The basic idea behind the witness tree method is to start with a ball at a large height and to construct a shallow tree with about $\log n$ nodes and depth about $\log \log n$ where nodes are the bins and an edge represents the two bin choices of a ball and argue that such a tree is unlikely to exist. While the construction as described may not necessarily produce a tree, we show later that the graph obtained has only a few extra edges in addition to a tree structure.

**Construction of the witness graph**: The root of the witness graph is the node with load $l + c$ where $c$ is a constant. For each of the top $l$ balls in this node, there must be an alternate bin choice. These alternate bin choices are set to be the children of the root node. The edges are labeled by the corresponding balls. Similarly we recurse for each of the $l$ children. For a parent node with load $x$, the $i^{th}$ ball from the top at height $x - i + 1$ must have had an alternate bin choice with load at least $x - i$; this is the $i^{th}$ child of the parent node. The $i^{th}$ child, which has load at least $l - i + c$, will be expanded down further to $l - i$ children corresponding to the alternate bin choices for its top $l - i$ balls. We continue this recursion as long as a node has load greater than $c$ and stop when it equals $c$ which corresponds to the leaf nodes. We also label each node (bin) with the set of lowest $c$ balls. These balls are distinct from the balls corresponding to the edges incident on this node. Clearly, this witness graph must be a subgraph of the original underlying graph on the bins.

It is possible that during this process, we run into cycles, that is, the child of some node may be an already existing node in the tree. For now, let us assume that this does not happen and the graph constructed is truly a tree. We will later argue that there cannot be too many cycle producing edges. Intuitively this is true because we are unlikely to run into cycles when a small number of $\log n$ nodes are explored as edges are chosen randomly.

Assuming that no cycles are found, after this process, we have a tree where each node with load $i + c$ has $i$ children with loads $i - 1 + c, i - 2 + c, \ldots, c$ respectively. Each edge corresponds to a distinct ball and each node is associated with $c$ additional distinct balls. Each edge indicates a distinct ball whose bin choices are exactly the end points of the edge. It is easy to check inductively that the total number of nodes in this tree is exactly $2^l$ (the recurrence relation for the number of nodes is $f(l) = f(l-1) + \ldots + f(0) + 1$ where $f(0) = 1$).

We now show that there cannot exist a set of $\log n$ nodes connected by a tree where each edge represents a distinct ball and further each node has a constant $c$ number of distinct additional balls.

---
[1] While the lower bound applies to the algorithm that always picks two random bin choices at most distance $\Delta$ apart when bins are arranged in a line, it is easy to generalize to any algorithm that chooses two bins for every ball that are distance $\Delta$ apart in expectation.

**Lemma 2.1** *The probability that there exists a set of $\Omega(\log n)$ bins connected by a tree where each edge represents a distinct ball and further each node has an additional set of c distinct balls is at most $1/n^{\Omega(1)}$*

**Proof**: We will bound the probability of existence of such a tree, with say $m$ nodes, by counting the number of such possible trees and multiplying by the probability that a given tree exists. The total number of different "shapes" (two shapes are the same if they are isomorphic) for a rooted tree on $m$ nodes is at most $4^m$ [20].

*Choosing bins and balls for nodes and edges:* For a given shape, the root bin can be chosen in $n$ ways. Once a node has been chosen, a given child can be chosen in $\Delta$ ways. Hence the total number of ways of choosing all the nodes is at most $n\Delta^{m-1}$. For each of the $m-1$ edges, the ball can be chosen in at most $n$ ways and the probability that it falls in this edge out of the total of $n\Delta/2$ choices is $2/n\Delta$. So, for a given shape, the total probability multiplied by the number of ways of choosing nodes and edges in the tree is at most $n\Delta^{m-1}n^{m-1}(2/n\Delta)^{m-1} = n2^{m-1}$.

*Choosing additional c balls per node:* In addition, for a given set of $m$ nodes, we need $c$ distinct balls to be associated with each node. This means that there are a total of $cm$ balls that are associated with nodes in the tree, that is, each of these balls is chooses some node in the tree. These $cm$ balls can be chosen in $\binom{n}{cm}$ ways. Since each node has $\Delta$ edges incident on it, the $m$ nodes of the tree have a total of $m\Delta$ edges and each of these $cm$ balls must choose one of these edges. This probability is at most $\binom{n}{cm}(\frac{m\Delta}{n\Delta})^{cm} \leq (\frac{en}{cm})^{cm}(\frac{m\Delta}{n\Delta})^{cm} = (e/c)^{cm}$.

Putting it all together, the total probability of finding such a tree is at most $4^m n 2^{m-1}(e/c)^{cm} \leq n[8(e/c)^c]^m$. The idea is to choose $c$ to be a large enough constant so that this probability is $1/n^{\Omega(1)}$; this can be achieved by setting $m$ to $\Omega(\log n)$. □

**Remark 2.1** *While the witness tree arguments in prior work [21, 29] do not look at all possible shapes of trees on m nodes, we use this approach as it extends easily to the witness tree analysis in the case when balls moves are allowed (Section 3).*

Since a witness tree with root of load $l + c$ has $2^l$ nodes, for $l = \log \log n + \Omega(1)$ the number of nodes becomes $\Omega(\log n)$; by Lemma 2.1, this is unlikely with high probability. So far we have assumed that the witness graph is a tree, ignoring the possibility of having cycles. We will show that the probability of finding many cycle-producing edges is very low. During the construction of the witness graph, an edge that leads to an existing node is called a cycle-producing edge (such edges are called *pruning* edges in past work [21, 29]). Let $p$ be the number of cycle-producing edges.

The next lemma shows that there cannot exist a witness graph with too many cycle-producing edges.

**Lemma 2.2** *The probability that there exists a set of $O(\log^2 n)$ bins satisfying the following conditions is at most $1/n^{\Omega(1)}$ for sufficiently large constants k and c.*

- *the bins are connected by a tree where each edge represents a distinct ball.*

- *there are additional $p = \frac{k \log n}{\log(\Delta/\log^4 n)}$ edges between these nodes representing distinct bins.*

- *and each node (bin) has an additional set of c distinct balls.*

**Proof**: As in proof of Lemma 2.1, we will bound the probability by counting the number of ways of choosing the balls, bins and edges and multiplying by the probability of each case. Let us consider a tree with $m$ nodes and $p$ additional edges. For a given shape of the tree, the additional $p$ edges can be chosen in at most $m^{2p}$.

For each of the $p$ additional edges, balls can be chosen in $n$ ways and the probability that the ball chooses the edge is $2/n\Delta$, giving a total probability of $n^p(2/n\Delta)^p = (2/\Delta)^p$.

As in proof of Lemma 2.1, the number of ways of choosing shapes for the tree, bins for the nodes of the tree and balls for the edges and $c$ balls for each node, multiplied by the probability of each case is at most $n[8(e/c)^c]^m$.

So the total probability of finding a witness graph with $m$ nodes and $p$ cycle-producing edges is at most $m^{2p}(2/\Delta)^p n[8(e/c)^c]^m = (2m^2/\Delta)^p n[2(e/c)^c]^m$.

If $p = \Omega(\frac{\log n}{\log(\Delta/\log^4 n)})$, then this probability is $1/n^{\Omega(1)}$. □

We are now ready to prove the main theorem.

**Proof of Theorem 2.2**: To take into account the possibility of cycle-producing edges, instead of starting with a node of load $l+c$, we start with a node of load $l+c+p$ where $p = k\frac{\log n}{\log(\Delta/\log^4 n)}$. At the first level, we expand to $p$ children corresponding to the top $p$ balls. The $i^{th}$ child of the root will have $l + p - i + c$. For this child, we treat $c' = p - i + c$ and expand as before till every leaf has load exactly $c'$, giving exactly $2^l$ nodes in this subtree (where nodes may be repeated). The total number of nodes, $M = 2^l p + 1 = p \log n + 1 \leq k \log^2 n + 1$ for $l = \log \log n$. The probability that there are more than $p$ cycle-producing edges is at most $1/n^{k-1}$. If there are less than $p$ cycle-producing edges, then during the construction of the witness graph, for at least one of the $p$ children, the construction below it is free of cycle-producing edges. This gives a witness tree where the root node has load at least $l + c$. By Lemma 2.1, the probability of this occurrence is at most $1/n^{\Omega(1)}$. So the probability of finding a witness graph whose root node has load $\log \log n + p + c$ is at most $1/n^{\Omega(1)}$. □

We also show that this result does not hold for non-regular graphs even if the minimum degree is $\Delta$. The counter-example is a complete bipartite graph with $n-\Delta$ nodes on one side and $\Delta$ on the other.

**Lemma 2.3** *There exist graphs with minimum degree $n^\epsilon$ for any constant $\epsilon < 1$ such that after insertion of $n$ balls by the same process as above, the maximum load of a bin is $\Omega(\frac{\log n}{\log \log n})$ with high probability of $1 - 1/n^{\Omega(1)}$.*

**Proof**: Consider the complete bipartite graph with $n-\Delta$ nodes on one side (left) and $\Delta$ on the other (right). We break the $n$ inserts into $\log n$ phases of $n/\log n$ inserts each. The essential idea is to show that for most of the early phases, after $i$ phases, each of the $\Delta$ nodes on the right side has load at least $i$ and at least $1/(4\log n)^i$ fraction of the left side vertices has load $i$ as long as $n/(4\log n)^{i+1} > \Delta \log n$. We prove this inductively. Assuming it is true at the end of the $i^{th}$ phase, in the first $n/(2\log n)$ inserts in the $(i+1)^{st}$ phase, at least $n/(4\log n)^{i+1}$ of them in expectation choose the left vertex to be of load at least $i$. This by assumption is at least $\Delta \log n$ and hence, if ties are broken to the right, then with high probability, all the $\Delta$ vertices on the right side would have been chosen and hence would have incurred a load of at least $i+1$. For the remaining $n/(2\log n)$ inserts of the phase, in expectation, at least $n/(4\log n)^{i+1}$ of the left vertices with load $i$ are chosen and they will all incur achieve a load of at least $i+1$. This goes on as long as $n/(4\log n)^i > \Delta \log n$ which means that the maximum load is at least $i = \frac{\log(n/(\Delta \log n))}{\log \log n}$. For $\Delta = n^\epsilon$, this is $\Omega(\frac{\log n}{\log \log n})$. All expectation results can be shown to hold with high probability using Chernoff bounds. □

We now show that when the number of balls $m$ is more than the number of bins $n$, the earlier results by Berenbrink *et al.* [3] do not carry over; even for graphs of degree $n^\epsilon$ where $\epsilon$ is constant, the maximum load can be $m/n + \omega(\log \log n)$.

**Lemma 2.4** *There exist $n^\epsilon$-regular graphs for any constant $\epsilon < 1$ such that after insertion of $m$ balls by picking edges at random, the maximum load of a bin is $m/n + \omega(\log \log n)$ with high probability of $1 - 1/n^{\Omega(1)}$ for sufficiently large $m$.*

**Proof**: The graph we use is a collection of $n^{1-\epsilon}$ cliques of size $n^\epsilon$. Some clique must get $m/n^{1-\epsilon} + \sqrt{m \ln n^{1-\epsilon}/n^{1-\epsilon}}$ balls when $m \geq n \ln n$ as this is like throwing $m$ balls into $n^{1-\epsilon}$ bins [27]. Hence some bin in this clique must get $m/n + \sqrt{m \ln n^{1-\epsilon}/n^{1+\epsilon}}$ balls. The result follows by setting $m = \Omega(n^{1+\epsilon})$. □

## 3 Balls and bins with moves

If moves are allowed during inserts of balls then the maximum load can be reduced further [24, 18]. On a complete graph, by performing at most $h$ moves per insert, we can maintain a maximum load of $O(\frac{\log \log n}{h \log(\log \log n/h)})$ [25]. In particular by performing up to $\log \log n$ moves we achieve a constant maximum load.

To understand how moves help in reducing load we view the balls and bins as nodes and edges of a graph which is a subgraph of the underlying graph; since multiple balls may choose the same pair of bins it is actually a multigraph. By making this graph directed, we could use the direction of an edge to indicate the choice of the bin among the two for placing the ball. The direction of each edge is chosen online by a certain procedure. The load of a vertex (bin) is equal to its in-degree. For each edge insertion, the two-choice algorithm directs the edge towards the vertex with the lower in-degree. During the ball insertion process, say $U$ is one of the vertices (bins) a ball chooses. Observe that if $VU$ is a directed edge, and if the load on $V$ is significantly lower, we could perform a move from $U$ to $V$, thus freeing up a position in $U$. Essentially, in terms of load, the new ball could be added to either $U$ or $V$, whichever has a lower load. This principle could be generalized to the case where there is a directed path from $V$ to $U$, and would result in performing moves and flipping the directions along all the edges on the path. If there is a directed sub-tree rooted at $U$, with all edges leading to the root, we could choose the least loaded vertex in this tree to incur the load of the new ball.

We now extend these bounds on maximum bin size with at most $h$ moves per insert to $\Delta$-regular graphs. We again use the same method as in section 2.2 but using a different witness tree. We use the same witness tree as described in [25]. Starting from a node with load $4l + c$, we get a witness tree with $l^{lh}$ nodes assuming no cycles are found during the construction. Further each node has a set of additional $c$ balls. So for $l = \Omega(\frac{\log \log n}{h \log(\log \log n/h)})$ the number of nodes becomes $\Omega(\log n)$.

Lemmas 2.1 and 2.2 continue to hold as they only refer to random choices of edges for balls which are same as before. So again there cannot be a witness tree with $\Omega(\log n)$ nodes or a witness graph with more than $p = k\frac{\log n}{\log(\Delta/\log^4 n)}$ cycle-producing edges. Again as before if we start with a node with load $4l + c + k\frac{\log n}{\log(\Delta/\log^4 n)}$ then at least one of the children of the root must be such that the witness graph construction under that node is free of cycle-producing edges, giving the following desired result.

**Theorem 3.1** *Given a $\Delta$-regular graph with $n$ nodes representing $n$ bins, if $n$ balls are thrown into the bins*

by choosing a random edge for each ball and performing at most $h$ moves as described above, then the maximum load is $O(\frac{\log \log n}{h \log(\log \log n/h)} + \frac{\log n}{\log(\Delta/\log^4 n)})$ with high probability of $1 - 1/n^{\Omega(1)}$.

## 4 Choosing bins in groups

Vöcking [29] observed that asymmetry helps in load balancing and showed that if ties are broken to the left when $n$ balls are inserted into $n$ bins with $d$ random choices per ball, the maximum load is $\frac{\log \log n}{d \ln \phi_d} + O(1)$ where $\phi_d$ approaches $\ln 2$ for large $d$.[2] This is a significant improvement over the $\frac{\log \log n}{\log d}$ bound. He also showed that the above bound is tight, that is, even if the $d$ bins are chosen from an arbitrary fixed distribution, the maximum load is at least $\frac{\log \log n}{d \ln \phi_d} - O(1)$ irrespective of the policy used to decide the bin out of the $d$ choices for a given ball.

We achieve a similar bound without making $d$ random bin choices as in Vöcking's result. Our algorithm makes two random accesses and chooses $d/2$ consecutive bins from each access. If the bins represent memory locations in a hash table then two burst accesses may be more efficient than $d$ random accesses. The maximum load of any bin is $\frac{\log \log n}{d \ln \phi_2} + O(1) = \frac{1.38 \log \log n}{d} + O(1)$.

Our algorithm works as follows. Group the $n$ bins into disjoint groups of $d/2$ consecutive bins each. Call these super-bins. Pick two super-bins at random and select the super-bin with the lesser total load (breaking ties to the left) where the total load of a super-bin is the sum of the loads of its $d/2$ bins. Then place the ball into the least-loaded of the $d/2$ bins in the super-bin.

**Theorem 4.1** *If $n$ balls are inserted into $n$ bins by the above algorithm, the maximum load of any bin is $\frac{\log \log n}{d \ln \phi_2} + O(1)$ with high probability. Instead of using two random groups of size $d/2$, if we use $c \geq 2$ groups of size $d/c$, then the maximum load of any bin is $\frac{\log \log n}{d \ln \phi_c} + O(1)$.*

Note that this approaches Vöcking's bound as $c$ becomes large. Moreover, our result implies that we can achieve constant load by making two burst accesses of $\log \log n$ bins each.

Furthermore, it also turns out that this partitioning into groups of size $d/2$ is essential in achieving this bound. It would seem that the bound of Theorem 4.1 would continue to hold even if all bins are not grouped into super-bins of size $d/2$, but instead the algorithm picks two random bins at least $d/2$ apart and $d/2$ consecutive bins starting from each of the chosen bins. Surprisingly this turns out to be false, i.e., the distribution becomes much worse if instead of choosing from well-aligned groups, the algorithm is allowed to choose two random disjoint sets of $d/2$ consecutive bins. For $d = \log \log n$, instead of getting a constant load, we get a maximum load of $\frac{\log \log n}{\log \log \log n}$.

**Theorem 4.2** *Instead of grouping the $n$ bins into aligned groups of size $d/2$, if we pick two disjoint random sets of $d/2$ consecutive bins, then the maximum load is $(1-o(1))(\frac{\log \log n}{\log d})$ with high probability at least $1 - o(1)$. As before, the algorithm picks the set with the smaller total load and places the ball into the least loaded bin in the selected set.*

Similarly it turns out that it is also important to choose the lesser loaded of the two chosen super-bins, that is, even with aligned bins, if the algorithm inserts into the least loaded bin in the two super-bins without considering the total load of the super-bins, then the maximum load of a bin becomes higher.

**Theorem 4.3** *If we place the ball into the least-loaded of the $d$ bins (that is, both the super-bins put together) in the above scheme without looking at the total loads of the each of the super-bins, then the maximum load of any bin is $(1 - o(1))(\frac{\log \log n}{\log d})$ with high probability at least $1 - o(1)$.*

We now prove Theorem 4.1 by analyzing super-bin loads using two-choice load-balancing results.

**Proof of Theorem 4.1**: Since each insert into a super-bin always goes into the least-loaded bin, the loads of any two bins in a super-bin differ by at most 1. Hence it is sufficient to analyze the maximum load of a super-bin. This is equivalent to throwing $n$ balls into $2n/d$ bins by choosing two bins at random, breaking ties to the left. Berenbrink *et al.* [3] extended Vöcking's result when the number of balls $m'$ is greater than the number of bins $n'$ showing that the maximum bin size is at most $\frac{m'}{n'} + \frac{\log \log n'}{d \ln \phi_d} + O(1)$. Hence the maximum load of a super-bin is $\frac{d}{2} + \frac{\log \log(2n/d)}{2 \ln \phi_2} + O(1)$. Thus the maximum load of any bin is at most $2 + \frac{\log \log n + O(1)}{d \ln \phi_2}$. Similarly if the super-bins are of size $d/c$, the maximum load of any bin is at most $\frac{\log \log n}{d \ln \phi_c} + O(1)$. □

We now prove Theorem 4.2. The intuition behind why the load increases when the groups are not aligned is the following: Let us artificially group the bins into aligned super-bins of size $d$ although this is not what the algorithm does. Let us estimate the fraction of bins with load at least $i + 1$, $p_{i+1}$ recursively. Earlier for a super-bin $I$ to grow from $i$ to $i + 1$, the second super-bin that was chosen must have at least $i$ balls. This is no more true if super-bins need not be aligned. Consider two adjacent aligned super-bins $I$ and $J$ with loads $i$ and $j = \lfloor i/d \rfloor d$ respectively. If $j = kd$, all bins in $I$ have load either $k$ or $k+1$ and all bins in $J$ have load $k$ (see Figure 1). If $i$ is not a multiple of $d$, then it is possible that the bin $b_I$ in $I$ next to $J$ has load $k$. If the algorithm chooses the set $I'$ of $d$ consecutive bins with the last bin $b_I$ in $I$ and the adjacent $d-1$ bins from $J$, then note that load of $I'$ is $j$. If the algorithm chooses to insert into super-bin $I'$, it could very well place the new ball in $b_I$, as all bins in $I'$ have equal load. This event happens if the other super-bin chosen had load at least $j$. The intuition is that since $j$ could be as low as $i - d$ and probabilities drop quickly, the probability that the $d$ bins next to $I$ have load $k$ each and that $I'$ is one of the super-bins chosen by the algorithm and the other super-bin chosen has load at least $j$ is much larger than $p_i^2$. Essentially the second super-bin need not have load $i$; it could have load $j$ that could be as small as $i - d + 1$. Informally speaking, the probability that the super-bin $J$ has load $j$ is $p_j$; the probability that the second super-bin chosen has load at least $j$ is also roughly $p_j$; probability that the super-bin $I$ has load $i$ is $p_i$. So the probability that the load of $I$ increases by one in a step could be about $p_i p_j p_j$, which is much lower than $p_i^2$, as $j$ can be as small as $i - d$. While this reasoning is not formal, to convert this into an inductive proof, we will need the neighboring super-bin of $J$ to have load $j - d$ and its neighboring super-bin to have load $j - 2d$ and so on.

---

[2] $\phi_d = \lim_{k \to \infty} \sqrt[k]{F_d(k)}$ where $F_d(k)$ is a generalization of Fibonacci sequence defined recursively as $F_d(k) = \sum_{i=1}^{d} F_d(k-i)$ and $F_d(k) = 0$ for $k \leq 0$ and $F_d(1) = 1$.

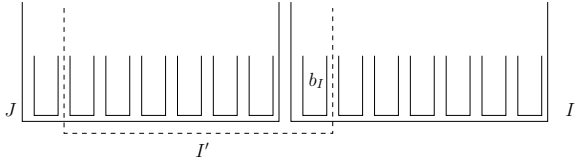

Figure 1: Unaligned super-bins: $I$ and $J$ are aligned, $I'$ is unaligned.

For ease of exposition, we will use groups of size $d$ instead of $d/2$. With a total of $n$ bins, we have $n/d$ super-bins. Let us group super-bins into $n/td$ blocks of $t$ consecutive super-bins each. (We will later set $t$ to be about $4d \log n$.)

**Definition 1** *A block of super-bins is a $k$-step if the first super-bin has all bins with load $k$, the next has all bins with load $k-1$, and so on and the last $t-k$ super-bins have all empty bins. Formally all bins in $l^{th}$ super-bin in the group have load $max(0, k-l)$ (Figure 2).*

We will track the number of $k$-steps in our bin configurations as inserts proceed in rounds of $n/t$ inserts. Let $q_r$ denote the fraction of blocks that are $r$-steps after $r$ rounds of $n/t$ inserts each. The following lemma provides a lower bound on $q_r$ recursively.

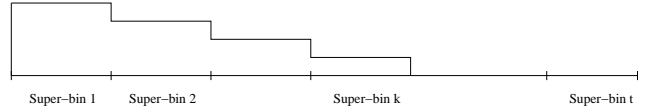

Figure 2: $k$-step: all bins in the $i^{th}$ super-bin have load $k - i$ if $i \geq k$ and 0 otherwise.

**Lemma 4.1** *Expected value of $q_r \geq (\frac{1}{10t^3d^2})^{rd} \cdot \prod_{i=0}^{r-1} q_i^d \cdot q_{r-1}$*

**Proof**: We will bound the fraction of $r-1$-steps that become $r$-steps after a round of $n/t$ inserts, only considering $r$-steps that are formed from $r-1$-steps by exactly a specific sequence of $rd$ insertion steps. Each of the bins in the $r^{th}$ through the first super-bin receive exactly one ball each in that order. Further the bins in the first super-bin receive balls in the left to right order. Precisely, the sequence of inserts is as follows:

1. The $r$-th super-bin is chosen as one of the two choices for insert and the insert ends up in this super-bin. This happens $d$ times making all bins in this super-bin to have load 1.

2. Similarly $d$ inserts happen in the $r-1$-st super-bin in the block causing all its bins to have load 2. Similarly for the $r-2$-nd, $r-3$-rd and so on till the second super-bin. After this, the block is almost an $r$-step except that the first super-bin needs to add an item to each of its bins.

3. We will insist that the last super-bin is handled exactly by the following process. Observe that all bins in the first and second super-bin have load $r-1$. Let $S_m$ denote the sequence of $d$ bins which are exactly the last $d - m$ bins from the first super-bin and the remaining $m$ from the second.

    (a) First $S_0$ (same as the first super-bin) is chosen for an insert and the new item goes into the first bin in $S_0$, which clearly is one of the least populated bins, as required for insert. Next, $S_1$ is selected for an insert and the new item goes into the first bin in $S_1$. And so on till $S_{d-1}$. In this way, all the bins in the first super-bin have load $r$ making the block an $r$-step.

Note that we require exactly these $rd$ inserts within the block in this sequence among the total of $n/t$ inserts. First let us find the number of $r-1$-steps that are chosen exactly $rd$ times in $n/t$ inserts. Probability that a given $r-1$-step gets chosen exactly $rd$ times is $\binom{n/t}{rd}(1/n)^{rd}(1 - (t+1)d/n)^{n/t - rd}$. This is $\geq (1/(trd)^{rd} e^{-2d})$. So the expected fraction of $r-1$-steps that get exactly the correct number of choices is $\geq (1/(10trd)^{rd})$.

Assuming an $r-1$-step gets chosen exactly $rd$ times in that order, let us now estimate the probability that all the $rd$ inserts go into the desired bins. The last $d$ inserts go to the right bins with probability at least $\frac{q_{r-1}}{td}$. This is because there are at least $\frac{q_{r-1}n}{td}$ unaligned-super-bin choices out of the total $n$ that have all bins of size at least $r-1$ and if the second choice is made from any of these, then it will clearly be larger than the choice in the $r-1$-step in consideration and so the insert can happen in the right bin in the $r-1$-step. Similarly the $i^{th}$ set of $d$ inserts go to the right bins with probability at least $\frac{q_{i-1}}{td}$.

Hence $q_r = (\frac{1}{10trd})^{rd} \cdot \prod_{i=0}^{r-1}(\frac{q_i}{td})^d \cdot q_{r-1} \geq (\frac{1}{10t^3d^2})^{rd} \cdot \prod_{i=0}^{r-1} q_i^d \cdot q_{r-1}$. While we assumed that ties are broken in our favor, the proof goes through even if we assume that the ties are broken randomly. $\square$

This can be converted into a high probability bound with probability at least $1 - 2/\log n$ (see Appendix 5.1), giving the slightly weaker recurrence $q_r \geq (\frac{1}{4})^d \cdot (\frac{1}{10t^3d^2})^{rd} \cdot \prod_{i=0}^{r-1} q_i^d \cdot q_{r-1}$ as long as the right hand side of this bound is at least $\frac{\log n}{n/(td)}$ and $t \geq 4d\log n$. Now we are ready to prove Theorem 4.2.

**Proof of Theorem 4.2**: Now that we have the recurrence on $q_r$, we show inductively that $q_r \geq 1/(40t^3d^2)^{(d+2)^r}$. Let $Q_r = \prod_{i=0}^{r} q_r$. Multiplying both sides of the recurrence by $Q_{r-1}$, we get that $Q_r \geq (\frac{1}{40t^3d^2})^{rd} \cdot (Q_{r-1})^{d+1} \cdot q_{r-1} \geq (\frac{1}{40t^3d^2})^{rd} \cdot (Q_{r-1})^{d+2}$. Inductively it is easy to check that $Q_r \geq (\frac{1}{10t^3d^2})^{(d+3)^r}$. The recurrence can be used as long as $q_r \geq \frac{\log n}{n/(td)}$ which holds as long as $Q_r \geq \frac{\log n}{n/(td)}$. This holds as long as $r \leq \log_{d+3} \log_{40t^3d^2} \frac{n}{td\log n} \leq \frac{\log\log \frac{n}{td\log n} - \log\log(t^3d)}{\log(d+3)}$. Setting $t = 4d\log n$, we get that $r = (1 - o(1))(\frac{\log\log n}{\log d})$. At this value of $r$, there are at least $\log n$ $r$-steps, completing the proof. The total proability of exceeding the bound is at most $O(r/\log n)$ which is $o(1)$. $\square$

Next we prove Theorem 4.3.

**Proof Sketch of Theorem 4.3**: All the bins in a super-bin with total load $i$ have load either exactly $\lfloor i/d \rfloor$ or $\lceil i/d \rceil$. The crucial idea is that it is no more true that in order for a super-bin $I$ with total count $i$, the probability of adding an extra ball is lower than $p_i^2$ where $p_i$ is the fraction of super-bins with load $i$. This is because if $i$ is just smaller than a multiple of $d/2$, then any bin with load $j$ where $j$ is the multiple of $d/2$ just less than $i$ could be chosen as one of the two choices along with $I$ and the next ball assignment could still go to $I$. This means that $p_{i+1} \approx p_i p_{\lfloor i/d \rfloor d}$. Analyzing this recursion gives the result. A formal proof can be obtained by using $\log n$ rounds of $n/\log n$ ball inserts as in the proof of Lemma 2.3. $\square$

# 5 Appendix

## 5.1 High probability version of Lemma 4.1

Here we convert the expected value of $q_r$ to a high probability bound for $q_r$.

The essential step is bounding the fraction of $r-1$-steps that receive exactly $rd$ inserts in the right order. While we proved earlier that the expected fraction is $\geq (1/(10trd)^{rd})$, we now show that the deviation from this mean is small and hence Chebyshev inequality can be applied to obtain high probability bound. Let $X_i$ denote the indicator variable for the event that the $i^{th}$ $(r-1)$-step gets the desired sequence of inserts.

**Lemma 5.1** $P(X_i X_j) \leq (1+\epsilon) P(X_i) P(X_j)$ where $\epsilon = 4d/t$.

**Proof:** $P(X_i) = P(X_j) = \binom{n/t}{rd}(1/n)^{rd}(1-(t+1)d/n)^{n/t-rd}$

$P(X_i X_j) \leq \binom{n/t}{rd}\binom{n/t-rd}{rd}(1/n)^{2rd}(1-2td/n)^{n/t-2rd}$

Comparing the terms, we get that

$$\frac{P(X_i X_j)}{P(X_i) P(X_j)} \leq \frac{\binom{n/t-rd}{rd}}{\binom{n/t}{rd}} \cdot \frac{(1-2td/n)^{n/t-2rd}}{(1-(t+1)d/n)^{2(n/t-rd)}}$$

After simplification, this is less than $(1+4d/t)$. □

**Lemma 5.2** For any random variable $R = \sum X_i$, if $P(X_i X_j) \leq (1+\epsilon) P(X_i) P(X_j)$, then the variance of $R$, $var(R) \leq \mu + \epsilon \mu^2$, where $\mu$ denotes $E[R]$.

**Proof:** For any random variable $R = \sum X_i$, variance of $R$ equals $E[(\sum X_i)^2] - E[\sum X_i]^2 \leq \sum E[X_i^2] + 2\epsilon \sum E[X_i]E[X_j] \leq \sum E[X_i] + \epsilon(\sum E[X_i])^2 = \mu + \epsilon \mu^2$ □

Let $\mu$ denote the expected number of $(r-1)$-steps that receive the desired $rd$ inserts. So the probability that less than $\mu/2$ get the desired $rd$ inserts is at most $var(R)/(\mu^2/4) \leq 1/\mu + \epsilon$.

Assuming $q_r \frac{n}{td} \geq \log n$ and $t \geq 4d \log n$, this probability is at most $2/\log n$. So with probability at least $(1-2/\log n)$, half $\mu$ will get the desired number of inserts. After that, by Chernoff bounds, assuming $\mu/2 \geq \log n$, we get that at least half of $(\prod_{i=0}^{r-1} \frac{q_{r-1}}{td})^d$ fraction will become $r$-steps, except for an exponentially low probability. Chernoff bound is applicable as long as the expected fraction of $r$-steps, $q_r \frac{n}{td}$ is at least $\log n$. So both the high probability bounds apply as long as $q_r \frac{n}{td} \geq \log n$, giving the slightly weaker recurrence $q_r \geq \frac{1}{4} \cdot (\frac{1}{10t^3 d^2})^{td} \cdot q_{r-1}^{td+1}$.